\documentclass[pra,reprint,nofootinbib,notitlepage,floatfix]{revtex4-1}
\usepackage[pdftex]{graphicx}
\usepackage{amsmath,amssymb,amstext}
\usepackage{mathrsfs}

\usepackage{multirow}

\usepackage{pdfsync}

\usepackage{comment}
\usepackage{bbold}
\usepackage{dsfont}
\usepackage{color}
\usepackage{hyperref}

\newcommand{\ket}[1]{\left|#1\right>} % Ket Dirac's notation %
\begin{document}
%TITLE PAGE
\title{Quantum to classical transitions in causal relations}

\author{Katja Ried$^{1,2,3,4\ast}$, Jean-Philippe W. MacLean$^{1,2\ast}$,  Robert W. Spekkens$^{3}$ and Kevin J. Resch$^{1,2}$}
\affiliation{$^1$Institute for Quantum Computing, University of Waterloo, Waterloo, Ontario, Canada, N2L 3G1}
\affiliation{$^2$Department of Physics \& Astronomy, University of Waterloo, Waterloo, Ontario, Canada, N2L 3G1}
\affiliation{$^3$Perimeter Institute for Theoretical Physics, Waterloo, Ontario, Canada, N2L 2Y5}
\affiliation{$^4$Institut f\"ur Theoretische Physik, Universit\"at Innsbruck, Technikerstraße 21a, 6020 Innsbruck, Austria\\
$^\ast$These authors contributed equally to this work.}

%\date{\today}

\begin{abstract}
 The landscape of causal relations that can hold among a set of systems in quantum
 theory is richer than in classical physics.  In particular, a pair of time-ordered
 systems can be related as cause and effect or as the effects of a common cause, and
 each of these causal mechanisms can be coherent or not.  Furthermore, one can
 combine these mechanisms in different ways: by probabilistically realizing either
 one or the other or by having both act simultaneously (termed a physical mixture).
 In the latter case, it is possible for the two mechanisms to be combined
 quantum-coherently.  Previous work has shown how to experimentally realize one
 example of each class of possible causal relations.  Here, we make a theoretical and
 experimental study of the transitions between these classes. 
 In particular, for each of the two distinct types of coherence that can exist in mixtures of common-cause and cause-effect relations---coherence in the individual causal pathways and coherence in the way the causal relations are combined---we determine how it degrades under noise and we confirm these expectations in a quantum-optical experiment.
%  in a quantum-optical . We develop  a quantum-optical experiment that studies the robustness of these two types of coherence under the influence of controlled noise.
% In particular, we explore the two distinct types of coherence that can exist in mixtures of common-cause and cause-effect relations: coherence in the individual causal pathways and coherence in the way the causal relations are combined. We develop  a quantum-optical experiment that studies the robustness of these two types of coherence under the influence of controlled noise.
\end{abstract}

  \maketitle

\section{Introduction}

 The study of quantum correlations that arise from entangled particles has played an
 important role in the advancement of quantum physics and the development of quantum
 technologies \cite{obrien_photonic_2009}. As such, the loss of coherence has been
 extensively studied in quantum information, from quantum error correction
 \cite{bennett_mixed_1996} to the investigation of the loss of entanglement
 \cite{yu_sudden_2009}, and also as a way to probe the quantum-classical boundary
 \cite{zurek_decoherence_2003,raimond_manipulating_2001}.  Causal relations are
 a powerful way of structuring our understanding of complex systems \cite{Pearl_book,
 Spirtes_book} and a more complete picture of any theory requires an understanding of
 the possible causal relations.  Yet the interplay between causality and quantum
 theory gives rise to long-standing puzzles: famously, Bell inequality violations
 show that quantum mechanics is not compatible with the conventional account of
 causation \cite{WoodSpekkens_2015}. At the same time, novel types of causal
 relations that can only hold between quantum systems have been shown to provide
 advantages in computing and information processing \cite{Hardy_2007,
   Chiribella_2012, AraujoEtAl_2014, ProcopioEtAl_2015, RiedEtAl_2015}.  However, the
 effect of decoherence on these nonclassical causal relations has not yet been
 explored. Assessing how robust non-classical causal
 relations are against decoherence and experimental noise is thus a necessary step
 before exploiting them as a resource in quantum information processing. 

\begin{figure}[t!]
 \centering
 \includegraphics[scale=1]{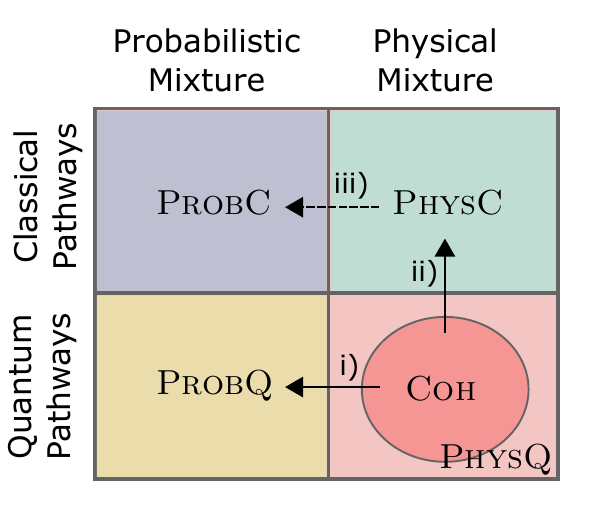} 
 \caption{\footnotesize{ \textbf{Classes of causal relations.} 
 A causal relation that is either cause-effect, common-cause, or a combination of
 both can be classified according to two criteria. The more familiar question is
 whether the mechanisms that realize the common-cause and cause-effect relations are
 themselves classical (C) or quantum (Q) that is, whether the common cause
 produces separable or entangled states and whether the cause-effect relation is
 a coherence-breaking or coherence-preserving channel. Another distinction is
 whether the two mechanisms are \emph{combined} in a probabilistic mixture (Prob),
 a physical mixture (Phys), or in a way that is intrinsically quantum (Coh). (Formal
 definitions are provided in the text.) These distinctions give
 rise to five categories of interest, as shown above.
%Note: there exists a sixth category, of classical mechanisms combined in a way that exhibits a quantum Berkson effect (obtained by dephasing only on B in our example of Coh), but we don't want to get into that. On the other hand, the quantum Berkson effect is a strict subset of general physical mixtures, which is why there aren't two more categories in this diagram (quantum Berkson intersecting probabilitic mixtures). 
 \textbf{(i)-(iii)} Our experiment explores the transition from the most restrictive %stringent
 category, \textsc{Coh}, to \textsc{ProbQ} and \textsc{PhysC} (solid arrows), as well
 as the classical transition from \textsc{PhysC} to \textsc{ProbC} (dashed arrow).}}
\label{fig:classification}
\end{figure}

 Another critical task in the development of quantum causality is the systematic
 classification of the possible causal relations between quantum systems.  We
 consider this problem in the case of two time-ordered quantum systems, where the
 possibilities are: a cause-effect relation, an unobserved common cause, or any
 combination of the two.  For such scenarios, we have previously proposed
 a classification of the possible causal relations, shown in
 Fig.~\ref{fig:classification}, which is based on two distinctions: whether the
 individual pathways (cause-effect or common-cause) are quantum or classical and
 whether they are combined probabilistically or in a more general way, termed
 a physical mixture. In Ref.~\cite{MacLeanEtAl_2016}, we investigated and
 experimentally realized particular examples of each class, including a fundamentally
 distinct type: a \emph{quantum-coherent mixture} of cause-effect and common-cause
 relations. 

 We now turn our attention to the transitions between these classes, in particular,
 the transitions that are induced by decoherence.  We explore the transition from
 fully  quantum-coherent mixtures of common-cause and cause-effect relations
 (\textsc{Coh}) to two other types of mixtures: 
 a probabilistic mixture of common-cause and cause-effect mechanisms wherein each mechanism exhibits quantum coherence (\textsc{ProbQ}) and a physical mixture 
 wherein both mechanisms are incoherent (\textsc{PhysC}).  
 %Note that a physical mixture of causal mechanisms is one wherein they act simultaneously~\cite{MacLeanEtAl_2016}.  a \emph{probabilistic mixture} (\textsc{ProbQ}) and a \emph{physical mixture} (\textsc{PhysC}), wherein the first case, each mechanism exhibits coherence individually and in the second, both are classical.

 These scenarios, which can only be represented using an extension of the standard quantum
 formalism, such as is provided in Refs.~\cite{AharonovEtAl_2009,Leifer_2006,ChiribellaEtAl_2009,Hardy_2012,
 Silva_2013,Oreshkov2012,Fitzsimons2013} and in particular
 Refs.~\cite{RiedEtAl_2015, MacLeanEtAl_2016}, exemplify the loss of two different
 types of coherence:  on the one hand, decoherence  in the individual mechanisms, and
 on the other, decoherence in the way the causal mechanisms are \emph{combined}.  The
 latter is a type of quantum-to-classical transition that has not been previously
 considered.
 %Studying this transition reveals how robust the quantumness of causal relations is against the addition of noise, which is particularly relevant for future experimental realizations. (You may be familiar with loss of coherence due to noise in entanglement or quantum channels. This paper is a bit like that, but cooler.)
 %For practical purposes, before using non-classical causal relations as a resource, one must verify how robust they are against noise.
 Moreover, by ranging over different families of causal maps instead of
 just isolated examples, we put the theoretical framework of
 Refs.~\cite{LeiferSpekkens_2013, RiedEtAl_2015, MacLeanEtAl_2016} to a more
 stringent test.
 
\section{Causal relations between two quantum systems}

\begin{figure}[t!]
 \centering
 \includegraphics[width=.8 \columnwidth]{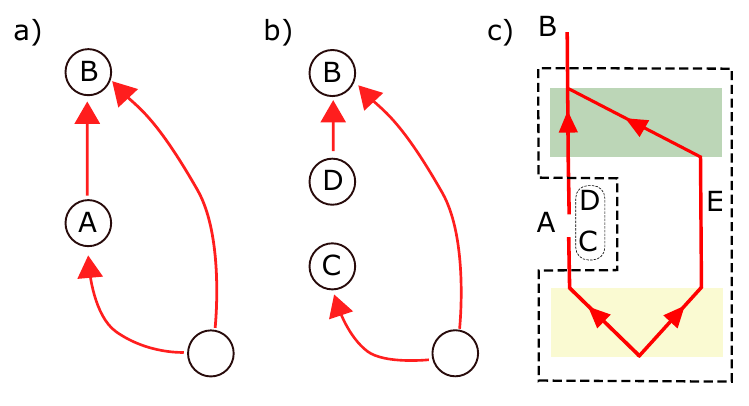} 
 \caption{\footnotesize{ {\bf General causal relation between two quantum systems.}
 (a) When the causal relations are represented as a directed acyclic graph, with
 arrows representing causal influences, one can distinguish the cause-effect pathway
 and the common-cause pathway. (b) In order to fully characterize the general causal
 relation between $A$ and $B$, the single system $A$ is replaced by two distinct
 systems: $C$ shares a common-cause relation with $B$, and $D$ has a cause-effect
 relation to $B$.  (c) The general causal relation between $A$ and $B$ is realized by
 the circuit fragment inside the dashed box, with $C$ and $D$ representing an output
 and input, respectively.  The label $E$ denotes the system that mediates the
 influence of the common cause on $B$.  }}
\label{fig:circuit}
\end{figure}

%Note: no classical examples in this paper!
%(Or maybe justify both the splitting and the form of the causal map in one go by pointing at a circuit.)
%In order to describe this causal relation completely, it is convenient to split $A$ into two versions: $C$, which has a purely common-cause relation with $B$, and $D$, which has a purely cause-effect relation with $B$. 
The  most general causal relation between two time-ordered quantum systems can be
realized by the circuit fragment shown in Fig.~\ref{fig:circuit}. Its
functionality, that is, the causal relation, is completely characterized by a trace-preserving,
completely positive map from the input of the circuit fragment, labeled $D$, to the
two outputs, $B$ and $C$: $\mathcal{E}_{CB|D}:\mathcal{L}(\mathcal{H}_D)\rightarrow
\mathcal{L}(\mathcal{H}_{CB})$,~\cite{RiedEtAl_2015,ChiribellaEtAl_2009,Hardy_2012,
Oreshkov2012,Silva_2013}, and which we call the causal map.

We here adopt the classification of causal maps proposed in
Ref.~\cite{MacLeanEtAl_2016}, which we review presently.  A causal map descibes
a purely cause-effect relation if it reduces to the form $\mathcal{E}_{CB|D}
= \mathcal{E}_{B|D} \otimes \rho_C$ and a purely common-cause relation if it takes
the form $\mathcal{E}_{CB|D} = \rho_{CB} {\rm Tr}_D$.  A causal map
$\mathcal{E}_{CB|D}$ is said to encode a {\em probabilistic mixture} of cause-effect
and common-cause relations if it  can be realized as follows: there is a hidden
classical control variable, $J$, which only influences $B$, such that for every value
of $J$,  either $B$ depends only on $D$ or $B$ depends only on its common cause with
$C$.  (See the Supplemental Information of Ref.~\cite{MacLeanEtAl_2016} for
a discussion of why this definition is apt.) Any causal map that cannot be cast in
this form is termed a \emph{physical mixture}. 

%FIGURE 2: CIRCUIT

 The first criterion for classifying causal relations is whether they are
quantum-coherent in the common-cause pathway (which relates $B$ to $C$) and in the
cause-effect pathway (relating $B$ to $D$).
A causal map is said to exhibit
\emph{quantumness in the common-cause pathway} if and only if there exists an
orthogonal basis of pure states on $D$, labeled by $d$ and denoted $\{ \Pi_D^d\}$, such that preparing each of
these states generates a state on $C$ and $B$, $\tau_{CB}^d\equiv
\mathcal{E}_{CB|D}(\Pi_D^d)$, which is entangled. The definition of quantumness in the
cause-effect pathway is closely analogous: given a measurement on $C$, it assesses
whether the correlations between $B$ and $D$ that arise for each measurement outcome
on $C$ exhibit entanglement. A state $\tau_{BD}^c$ that encodes these correlations
can be constructed as follows: let $\Pi_C^c$ denote the operator associated with
a measurement outcome $c$, let $\tau_{CBD}$ denote the state that is
\emph{Choi isomorphic}~\cite{Choi1975} to $\mathcal{E}_{CB|D}$, and take
$\tau_{BD}^c\equiv \frac{1}{P(c)} {\rm Tr}_C \left(\Pi^c_C \tau_{CBD}\right)$, where
$P(c)={\rm Tr} \left(\Pi^c_C \tau_{CBD}\right)$ ensures unit trace. 
%To formalize this, let $\Pi^c$ be a rank-one projector associated with an outcome $c$ in the measurment, let 
%consider a measurement on $C$, whose outcomes $c$ are associated with rank-one projectors $\Pi^c$, let 
%$\mathcal{E}_{B|D}^c\equiv {\rm Tr}_C (\Pi^c_C \mathcal{E}_{CB|D})$ denote the map from $D$ to $B$ associated with that outcome and let  $\tau_{BD}^c $ denote the normalized Choi state that is isomorphic to that map. 
A causal map is \emph{quantum in the cause-effect pathway} if and only if  there exists a rank-one projective measurement on $C$ such that, for all outcomes $c$, the states $\tau_{BD}^c$  are entangled. 
In order to quantify the entanglement in the induced states  of the form $\tau^z_{XY}$, we compute their negativity~\cite{VidalWerner_2002},
\begin{equation}
\mathcal{N}^z_{XY}\equiv\frac{1}{2}({\rm Tr}|T_Y(\tau^z_{XY})|-1),
  \label{eq:negativity} 
\end{equation} 
where %$X$ and $Y$ are the two variables under consideration for a given preperation or
%measurement labelled $z$ and 
$T_Y(\cdot)$ denotes transposition on $Y$.

The second criterion for classifying causal relations is whether the common-cause and
cause-effect mechanisms are \emph{combined} as a probabilistic mixture or as
a physical mixture.  A special instance of the latter class is a quantum-coherent
mixture (the definition of which will be provided momentarily).  In order to
formalize and detect these distinctions, we note the following: if $B$ was either
purely cause-effect related to $D$ or purely common-cause related to $C$, then
acquiring new information about $B$ either leads one to update one's information
about $D$ or it leads one to update one's information about $C$, respectively, but it
does not give rise to correlations between $C$ and $D$. If such correlations are
observed, then, they herald a mixture of common-cause and cause-effect mechanisms.
Moreover the strength of the induced correlations can distinguish between
probabilistic and physical mixtures ~\cite{MacLeanEtAl_2016}. 

In order to assess these correlations, we define an induced state $\tau_{CD}^b$
similarly to $\tau_{BD}^c$ above: given a measurement on $B$ with outcomes $b$
associated with operators $\Pi^b$, let $\tau_{CD}^b\equiv \frac{1}{P(b)} {\rm Tr}_B
\left(\Pi^b_B \tau_{CBD}\right)$. [The normalization factor $P(b)$ is the probability
of obtaining outcome $b$ if one inputs the maximally mixed state on $D$ into the
causal map.] We apply two metrics to the induced state.

 In the first, we test whether two qubits are related by a probabilistic mixture of
 common-cause and cause-effect mechanisms. We consider a measurement of $\sigma_z$ on
 $B$, whose outcomes we label $b=\pm1$, introduce the covariance of the outcomes of
 Pauli measurements in state $\tau_{CD}^b$,
\begin{align}
{\rm cov}\left(\tau_{CD}^b\right) \equiv& 
%{\rm Tr} \left[ \sigma^x_C \otimes \sigma^y_D \tau_{CD}^b \right]  \\ \nonumber
%&- {\rm Tr} \left[ \sigma^x_C \otimes \mathbb{1}_D \tau_{CD}^b \right] \cdot 
%{\rm Tr} \left[ \mathbb{1}_C \otimes \sigma^y_D \tau_{CD}^b \right],
\langle \sigma^x_C \otimes \sigma^y_D \rangle- %_{ \tau_{CD}^b} - 
%\langle \sigma^x_C \otimes \id_D \rangle_{ \tau_{CD}^b} \langle \id_C \otimes \sigma^y_D \rangle_{ \tau_{CD}^b}
\langle \sigma^x_C \rangle %_{ \tau_{CD}^b} 
\langle \sigma^y_D \rangle %_{ \tau_{CD}^b}
\end{align}
and define the witness
\begin{equation}
\mathcal{C}_{CD}\equiv 2 \sum_{b=\pm1} b P(b)^2 {\rm cov}(\tau_{CD}^b).
\label{eq:CCD}
\end{equation}
It was shown in Ref.~\cite{MacLeanEtAl_2016} that $\mathcal{C}_{CD}=0$ for all probabilistic mixtures, hence non-zero values herald physical mixtures. 

The second metric tests whether the cause-effect and common-cause mechanisms are
combined coherently.  If there exists a rank-one projective measurement on $B$ such
that the states $\tau_{CD}^b$ are entangled for all possible outcomes $b$, then the
causal map is said to exhibit a \emph{quantum Berkson effect}: a quantum version of
Berkson's effect, whereby conditioning on a variable induces correlations between its
causal parents which are otherwise uncorrelated. Again, we quantify this in terms of
negativity, Eq.~\eqref{eq:negativity}. Following the proposal of
Ref.~\cite{MacLeanEtAl_2016}, a causal map is termed a \emph{quantum-coherent
mixture} of common-cause and cause-effect mechanisms if and only if (a) it exhibits
a quantum Berkson effect and (b) it is not entanglement breaking in both the common-cause and
cause-effect pathways.

\begin{figure*}[th]
  \centering
    \includegraphics[scale=1]{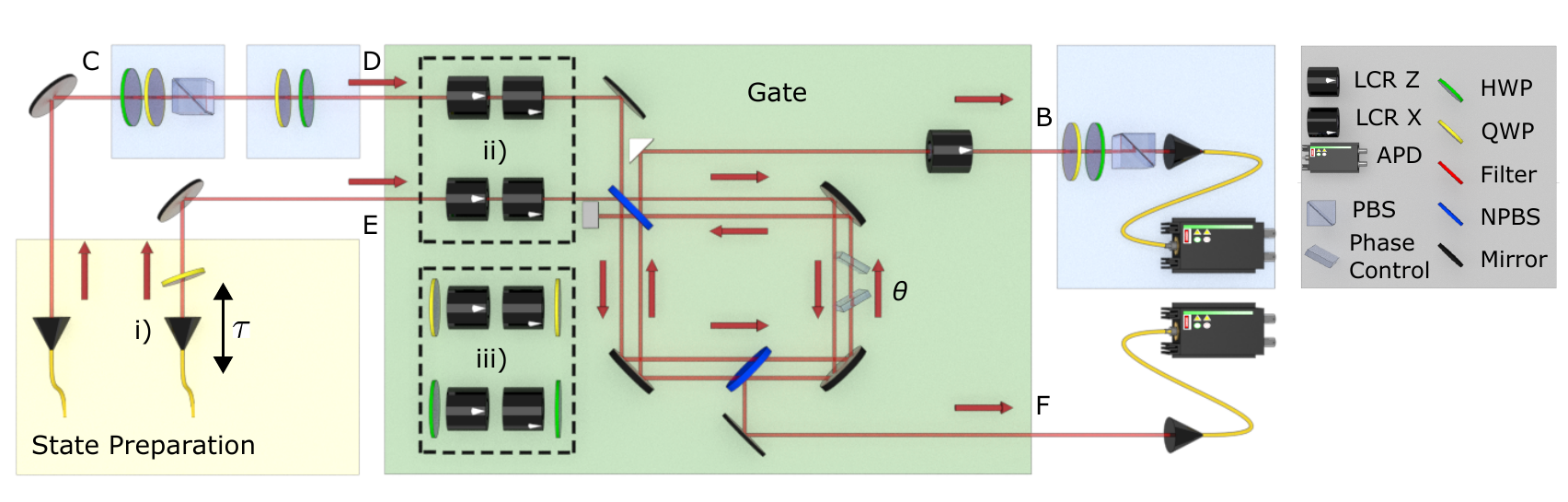} 
    \caption{\footnotesize{ {\bf Optical implementation for exploring causal relations.} 
     The initial preparation (yellow) targets the maximally entangled state
    $\ket{\Phi^+}$ from Eq.~\eqref{eq:phiplus}. The gate (green) allows the implementation of different ways of combining a common-cause relation (between $C$ and $B$) with a cause-effect relation (between $D$ and $B$). Measurements of $C$ and $B$ and repreparations of $D$ (blue) are used to characterize the causal relation. 
    %    combinations of common-cause and cause-effect relations of two quantum variables, $C$ (common-cause) and $D$ (cause-effect), to a third variable $B$.  
    %One photon is measured at $C$ and reprepared at $D$; then both are sent through the gate and one photon is measured at $B$. 
    %{\bf b)} Experimental setup including polarization entangled photon source and the partial swap gate which implements the unitary $U(\theta)=\mathrm{cos}(\theta/2)\mathds{1} +i\mathrm{sin}(\theta/2)\mathrm{SWAP}.$ 
    %The phase $\theta$ is controlled by tilting the angle of glass plates in the Sagnac     interferometer. 
    We explore the transitions \textbf{(i)} \textsc{Coh} to \textsc{ProbQ},  where we
    fix the dephasing probability ($p=0$) and phase ($\theta=\pi/2$) and vary the
    delay $\tau$ of the photon at $E$; \textbf{(ii)} \textsc{Coh} to \textsc{PhysC},
    where we fix the dephasing directions along $(\hat x,\hat y,\hat z)$ and vary the
    phase $\theta$ (by tilting the glass plates in the interferometer) and the
    dephasing probability $p$ (by switching the LCRs on with probability $p$);
    \textbf{(iii)} \textsc{PhysC} to \textsc{ProbC}, where we fix the dephasing
    probability ($p=1$) and phase ($\theta=\pi/2$) while interpolating from $(\hat
    n_E,\hat n_D,\hat n_B)=(\hat z,\hat z,\hat z)$ to $(\hat x,\hat y,\hat z)$ using
    a single-parameter family of gates (see Eq.~\ref{eq:eta}).  Notation for optical elements: half-wave
    plate (HWP), quarter-wave plate (QWP), liquid-crystal retarder (LCR), polarizing
  beam splitter (PBS), non-polarizing beam splitter (NPBS), avalanche photo diode
(APD).  }} \label{fig:setup} \end{figure*}

\section{Realizing families of causal maps}

 In Ref.~\cite{MacLeanEtAl_2016}, we used a circuit similar to Fig.~\ref{fig:setup}
 in order to realize four particular causal maps that constitute paradigmatic
 examples of the classes introduced in Fig.~\ref{fig:classification}.  Here we
 introduce a modified set-up in order to observe the transitions shown
 in Fig.~\ref{fig:classification}: \textsc{Coh} to \textsc{ProbQ}, \textsc{Coh} to
 \textsc{PhysC}, and \textsc{PhysC} to \textsc{ProbC}.

 The initial state $\rho_{CE}$ in all cases is the maximally entangled state
\begin{equation}
\ket{\Phi^+}  \equiv\frac{1}{\sqrt{2}}(\ket{HH}+\ket{VV}),
\label{eq:phiplus}
\end{equation}
 where $\ket{H}$, $\ket{V}$ denote respectively the horizontal and vertical
 polarization states.
 %eigenstates of the Pauli operator $\sigma_z$. 
 The entanglement in the initial state provides the
 common cause relation in the experiment and can be removed with full dephasing noise. 
 %The heart of the circuit is the
 %The circuit is composed of a quantum logic gate, called a partial swap, which consists of 
 Between $DE$ and $BF$ we apply the partial swap gate, which realizes a family of unitaries parametrized by the parameter $\theta$, 
 \begin{align}
    \begin{split}
    U_{BF|DE}=
   \cos\frac{\theta}{2}  \mathds{1}_{B|D} \otimes \mathds{1}_{F|E}
    + i \sin \frac{\theta}{2}  \mathds{1}_{B|E} \otimes \mathds{1}_{F|D},
  \end{split}\label{eq:partialswap}
  \end{align}
 where $\mathds{1}_{Y|X}$ denotes the identity operator with input space $X$ and output space $Y$.
 % from the variables $X$ to $Y$.  
 When $\theta=0$, the unitary reduces to the two-qubit identity operator,
 whereas when $\theta=\pi$, it reduces to the swap operator.  When $\theta=\pi/2$,
 this gate reduces to the square root of swap or {\em partial swap}
 operator \cite{CernochEtAl_2008,MacLeanEtAl_2016}.  When this partial swap gate is combined with
 the initial state on $CE$ described above, it was shown in Ref.~\cite{MacLeanEtAl_2016} that it realizes our paradigm example of \textsc{Coh}: a causal map
 that is quantum in the common-cause and cause-effect pathways and furthermore exhibits a quantum
 Berkson effect. 

 We first study a transition from our example of \textsc{Coh} to an example of
 \textsc{ProbQ},
 %\color{cyan} [Shouldn't we describe this example before describing the family that interpolates between the two extremes?]\color{black},
  a causal map that is a probabilistic mixture of common-cause and
 cause-effect relations, which are individually quantum. 
 To realize this transition, we introduce a delay $\tau$ between the two photons entering the
 interferometer.
 %that would normally realize the partial swap. This 
 In Ref.~\cite{MacLeanEtAl_2016}, it was shown that the causal map went from
 \textsc{Coh} to \textsc{ProbQ} as the delay $\tau$ was increased beyond the
 coherence length $\tau_{coh}$ of the photons. 
 We can exploit this to follow the evolution across this transition for intermediate
 values of $\tau$.
 %gradually reduces their indistinguishability and converts the gate
 %%from $DE$ to $BF$ 
 %to an equal probabilistic mixture of the two-qubit identity and swap gates. 
 %During this transition, 
 The gate is then modelled as
 \begin{align} 
   \begin{split}
   \mathcal{E}_{BF|DE}&(\cdot)=
   q U_{BF|DE} (\cdot) U_{BF|DE}^\dagger  \\ 
   &+ (1-q) \Big[ \tfrac{1}{2} \mathds{1}_{B|D} \otimes \mathds{1}_{F|E} (\cdot) \mathds{1}_{B|D} \otimes \mathds{1}_{F|E} \\ 
   &\hspace{1.3cm} + \tfrac{1}{2} \mathds{1}_{B|E} \otimes
   \mathds{1}_{F|D} (\cdot) \mathds{1}_{B|E} \otimes \mathds{1}_{F|D}\Big],
 \end{split}
 \end{align}
 where the factor $q=\exp (\tau^2/2\tau_{coh}^2)$ expresses the amount of
 temporal overlap between two pulses, as a function of the delay $\tau$ between them,
 assuming two Gaussian pulses with a (RMS) coherence length of $\tau_{coh}$.
 %the overlap in time of the two photons with pulse-width $\sigma_t$ as a function of the delay $\tau$ between the two pulses. 
 %As the temporal overlap between the pulses decreases, $q$ decreases from 1 to 0 and we range from our example of 
 Starting from our example of \textsc{Coh} ($q=1$),
 as the delay increases, the overlap of the photons decreases and we realize
 our paradigm example of \textsc{ProbQ}  when $q=0$. 

 %For the second quantum-classical transition, \textsc{Coh} to \textsc{PhysC},
% In order to transition from \texthis into examples in which the two mechanisms are 
% classical, one can add total 
 For the second transition, we apply dephasing channels on $D$, $E$ and $B$. The generic
 single-qubit dephasing channel is specified by the basis in which the dephasing
 occurs, represented by a Bloch vector $\hat n$, and the dephasing probability,
 denoted $p$: 
 \begin{align} \Delta(\hat{n},p)(\rho)\equiv(1-p/2)\rho+p/2\left([\hat
   n\cdot \vec{\sigma}]\rho[\hat n\cdot \vec{\sigma}]\right). \label{eq:dephase}
 \end{align}
 Complete dephasing ($p=1$) along $\hat y$ on $D$, $\hat x$ on $E$ and $\hat z$ on $B$  realizes our paradigm example of \textsc{PhysC}.   In this case, the gate is
 \small
  \begin{align}
    \label{eq:gatephysC}
      &\mathcal{E}_{BF|DE}(\cdot)\equiv\\\nonumber
      &(\Delta^B(\hat z,1)\otimes\mathcal{I}_F)\left(U_{BF|DE}\left\{ [
      \Delta^D(\hat y,1)\otimes\Delta^E(\hat x,1)](\cdot)\right\}
      U_{BF|DE}^{\dag}\right),
    \end{align}
  \normalsize
  where $\mathcal{I}_F$ is the identity map on $F$.
 The %second quantum to classical transition, from \textsc{Coh} to \textsc{PhysC},
 transition from \textsc{Coh} to \textsc{PhysC} is then realized by fixing the direction
 of dephasing ($\hat y$ on $D$, $\hat x$ on $E$ and $\hat z$ on $B$) but ranging over
 the dephasing probability, $p$. 

 Finally, note that complete dephasing  ($p=1$) along $\hat z$ on each of $D$, $E$
 and $B$  generates our paradigm example of \textsc{ProbC}.  Here, the gate is
 \small
 \begin{align}\label{eq:gateprobC}
      &\mathcal{E}_{BF|DE}(\cdot)\equiv\\ \nonumber
      &(\Delta^B(\hat z,1)\otimes\mathcal{I}_F)\left(U_{BF|DE}\left\{ [
      \Delta^D(\hat z,1) \otimes\Delta^E(\hat z,1)](\cdot)\right\}
      U_{BF|DE}^{\dag}\right).
 \end{align}
 \normalsize
 We can therefore realize a one-parameter family of causal maps that interpolate
 between our paradigm examples of  \textsc{PhysC} and \textsc{ProbC} by rotating  the
 bases in which $D$, $E$, and $B$ are dephased.  The directions along which each of
 these systems is dephased, with full dephasing strength ($p=1$), is parameterized by
 $\eta$, and given as:
 \begin{align}
   \begin{split}
   \hat  n_E =&\cos{2\eta}~\hat
  x+\sin{2\eta}~\hat z,\\ \hat n_D=&\cos{2\eta}\sin 2\eta~\hat x-\cos
  2\eta~\hat y+\sin^2{2\eta}~\hat z, \\ \hat n_B=&\hat z.  
   \end{split}
   \label{eq:eta}
\end{align}
As one varies between $\eta=0$ and $\eta=\pi/4$, one transitions between full dephasing along the triple of axes $(\hat n_E,\hat n_D,\hat n_B)=(\hat x,\hat y, \hat z)$ to full dephasing along the triple of axes $(\hat z, \hat z, \hat z)$ and hence between our paradigm examples of \textsc{PhysC} and of \textsc{ProbC}. \color{black}

%These unit vectors describe the directions along which full dephasing ($p=1$) is applied, when transitioning from \textsc{PhysC} ($\eta=0$) to \textsc{ProbC} ($\eta=\pi/4$).

%SECTION: EXPERIMENT 
\section{Experiment}

 The experimental set-up is shown in Fig.~\ref{fig:setup}. We aim to produce
 polarization-entangled photons at 790 nm in the state $\ket{\Phi^+}$. 
  %The experimental set-up is shown in Fig.~\ref{fig:setup}.  We use a Ti:Sapphire
  %laser, with a  repetition rate of 80 MHz, a central wavelength of 790nm, and an
  %average power of 2.65 W. The laser light is frequency doubled in a 2-mm thick
  %bismuth-borate (BiBO) crystal, creating a pulses at 395nm (0.65 W average power)
  %with 1 nm FWHM.  In a pair of 1~mm $\beta$-barium-borate (BBO) crystals with
  %orthogonal orientations, the pump is focused for type-I spontaneous parametric
  %downconversion.  Polarization entangled photon pairs at 790 nm are produced in the
  %target state $\ket{\Phi^+}$. Bandpass filters reduce background noise and
  %inteference filters set the photon bandwidths to 3 nm.  Additional compensation
  %crystals counteract the effects of temporal and spatial walkoff in the crystal.
  %\cite{lavoie_experimental_2009}.  The photons are coupled into single mode fibres
  %where the polarization is set with polarization controllers and the phase of
  %the entangled state is tuned by tilting a quarter-wave plate (QWP) at the output of
  %one of the fibres.
 The polarization of one photon is measured at $C$ using a half-wave plate (HWP),
 quarter-wave plate (QWP), and polarizing beam splitter (PBS) in sequence, after
 which the photon is reprepared in another polarization state at $D$.  At $D$ and
 $E$, both photons are then sent towards the quantum circuit, which uses a folded
 displaced Sagnac interferometer \cite{MacLeanEtAl_2016}, with two 50/50 beam
 splitters and two NBK-7 glass windows that are counter rotated in order to change
 the phase $\theta$ with minimal beam deflection. The polarization of the photon at
 $B$ is measured and detected in coincidence with the one at $F$. The photon (RMS)
 coherence length, $\tau_{coh}=189$ fs, is estimated using a Hong-Ou Mandel
 interference measurement at the first beam splitter assuming transform-limited
 Gaussian pulses. 
 % FWHM of $\sigma_t=371fs$. 
 
 The implementation of the partial swap $U_{BF|DE}$ of Eq.~\eqref{eq:partialswap} in an
 optical circuit follows \cite{CernochEtAl_2008, MacLeanEtAl_2016}.
 When two indistinguishable photons arrive at the first beam splitter, they will
 bunch if their polarization state lies in the symmetric subspace, and they will
 anti-bunch if their polarization state lies in the anti-symmetric subspace. If they
 bunch, only the clockwise path in Fig.~\ref{fig:setup}, which introduces no phase
 shift, can lead to a coincidence at $B$ and $F$.  However, if they anti-bunch, the
 photon that travels along the counter-clockwise path will acquire an extra phase
 $\theta$, while the other, on the clockwise path, acquires no extra phase. Since the
 photons are indistinguishable, these two paths are coherently recombined and as
 a result the gate applies a phase difference $\theta$ between the symmetric and
 anti-symmetric projections of the photon state, leading to the partial swap unitary
 of Eq.~\eqref{eq:partialswap}.  Further details on the single photon source and Sagnac
 interferometer are provided in Ref.~\cite{MacLeanEtAl_2016}. 

 %{[\color{red} Same as previous experiment]} The visibility of the Sagnac
 %interferometer without background subtraction is $\left( 93.6\pm0.1 \right)\%$.
 %This is measured by blocking one input ($D$ or $E$) to the gate and measuring the
 %number of photons at the output $B$ as a function of the window angles in the Sagnac
 %interferometer (see supplementary material).  For the partial swap to be effective,
 %photon pairs in modes $D$ and $E$ must undergo two-photon quantum interference on
 %a beamsplitter prior to entering the gate.  Hong-Ou-Mandel interference between
 %photons input at $D$ and $E$ is measured at the first beam splitter using
 %a translation stage on input $E$.  A dip in visibility of $(92.5\pm1.7)\%$ is
 %achieved.  %The delay is set to a value correponding to the %centre of the HOM dip
 %to maximize the two photon interference.

  The dephasing channels before and after the Sagnac interferometer are implemented
  using variable liquid crystal retarders (LCR), which apply a voltage-dependent
  birefringence, introducing a phase shift of $0$ or $\pi$.  
 % Probabilistic dephasing is achieved by switching them on and off at random at
 % a rate of $10~Hz$, with a probability $p$ of implementing a $\pi$ phase shift
 % during each interval.  
  Each axis for dephasing can be chosen independently. Dephasing along the $\hat z$
  direction of the Bloch sphere is achieved by putting the LCR axis at $0^\circ$,
  dephasing along the $\hat x$ direction, by putting the LCR at $45^\circ$, and in
  the $\hat y$ direction, by simultaneously switching on two LCRs, one at $0^\circ$
  and one at $45^\circ$.  Dephasing in the rotated basis is achieved by setting two
  QWPs on either side of an LCR at $0^\circ$ and two HWPs on either side of an LCR at
  $45^\circ$.  The transition from full dephasing along the triple of axes $(\hat
  n_E,\hat n_D,\hat n_B)=(\hat x,\hat y, \hat z)$ to full dephasing along the triple
  of axes $(\hat z, \hat z, \hat z)$ that is described in Eq.~\eqref{eq:eta}
  %  from $(\hat n_E,\hat n_D,\hat n_B)=(\hat x,\hat y, \hat z)$ dephasing to $(\hat z, \hat z, \hat z)$ at $p=1$ 
  %in Fig.~\ref{fig:rotated witness} 
  is therefore 
  implemented with the following transformations on $E$, $D$, and $B$, 
  \begin{align*}
    & E: U_{\mathrm{HWP}}(\eta/2)\Delta (\hat x,p=1)U_{\mathrm{HWP}}(\eta/2)\\
    & D: U_{\mathrm{QWP}}(3/4\pi-\eta)\Delta (\hat z,p=1)U_{\mathrm{QWP}}(\pi/4-\eta)\\
    & B: \Delta (\hat z,p=1),
    \label{}
  \end{align*}
  where $U_{\mathrm{HWP}}(\eta)$ and $U_{\mathrm{QWP}}(\eta)$ specify the unitaries
  for a HWP and QWP, respectively, with the fast axis at an angle $\eta$, while $\Delta
  (\hat x,p=1)$ and $\Delta (\hat z,p=1)$ are the completely dephasing channels
  implemented with LCRs.

 \section{Results} 

%part 1: test witness of physical mixture

%noise and theoretical curves that take it into account
\begin{figure}[t]
   \centering
   \includegraphics[scale=1]{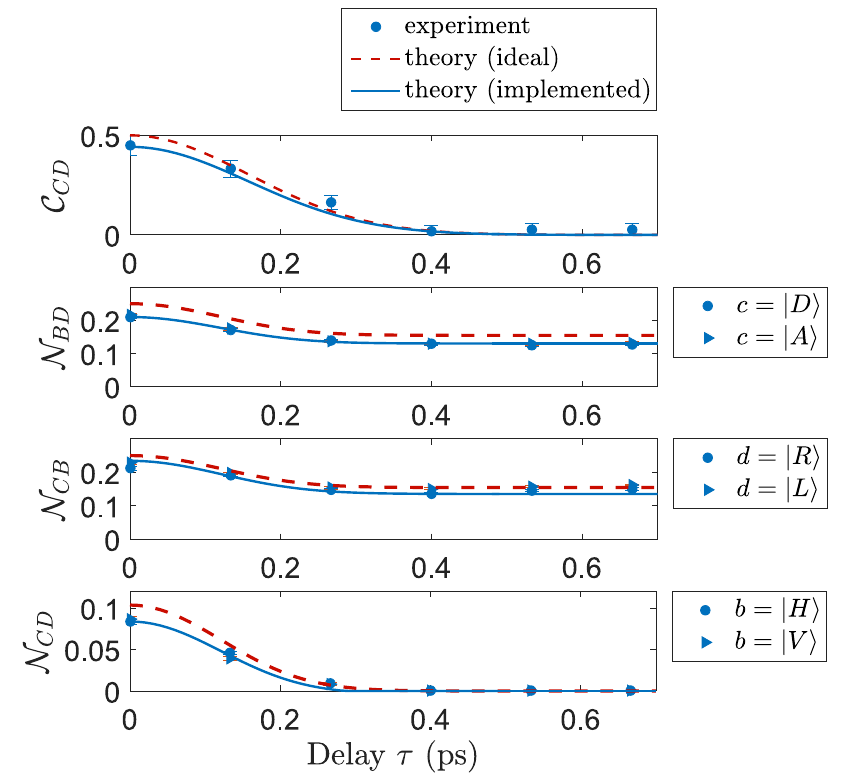}
   \caption{\textbf{Transition from Coh to ProbQ.} As the delay $\tau$ between
   photons is increased, quantum interference at the gate is gradually lost. We
   observe: (a) the witness of physical mixture $\mathcal{C}_{CD}$; %which is zero for all probabilistic mixtures; 
   (b) the negativity $\mathcal{N}_{BD}$ under post-selection on eigenstates of $\sigma_X$ on $C$; %which witnesses quantumness in the cause-effect pathway; 
   (c) the negativity
   $\mathcal{N}_{CB}$ under pre-selection on eigenstates of $\sigma_Y$ on $D$; %which witnesses quantumness in the  common-cause pathway; 
   and (d) the negativity
   $\mathcal{N}_{CD}$ under post-selection on eigenstates of $\sigma_Z$ on $B$.
   %which   witnesses %is a necessary condition for quantumness in the way common-cause and cause-effect paths are combined.  
   (b),(c) Nonzero values of $\mathcal{N}_{BD}$ and $\mathcal{N}_{CB}$ herald
   quantumness in the common-cause and cause-effect pathways, which is preserved
   throughout the transition, while (a),(d) the quantum Berkson effect is supressed before
   the witness of physical mixture reaches zero. For intermediate
   values, between $\tau=0.27$ ps ($q=0.23$) and $\tau=0.39$ ps ($q=0.05$), we realize
   a causal map that belongs to \textsc{PhysQ}, but observe no evidence that it
   belongs to \textsc{Coh}. For larger delays, there is no evidence of physical
   mixture ($\mathcal{C}_{CD}=0$), making the data consistent with \textsc{ProbQ}. Solid
 lines represent theoretical predictions in the ideal case (red) and taking into
 account the experimental implementation (blue), as described in the text.}
 \label{fig:coh2probq}
 \end{figure}

 In Fig.~\ref{fig:coh2probq}, we observe signatures of the 
 transition from \textsc{Coh} to \textsc{ProbQ}, which is
 generated by delaying one photon relative to the other before the gate.  The
 indicators of quantumness in the common-cause and cause-effect pathways,
 $\mathcal{N}_{CB}$ and $\mathcal{N}_{BD}$, remain non-zero throughout the
 transition, whereas
%even in the limit of a probabilitistic mixture. 
 the witnesses of physical mixture, $\mathcal{C}_{CD}$, and of the quantum Berkson
 effect, $\mathcal{N}_{CD}$, fall to zero as the delay increases, since the photons become
 distinguishable and coherence in the gate is lost.
 Thus, although the individual common-cause and cause-effect mechanisms remain
 quantum, we observe a loss of coherence in the way in which they are
 combined.  This exemplifies the transition from
 %we consider the family of causal maps that range from
 \textsc{Coh} to \textsc{ProbQ}.

% on the other hand, falls to zero as overlap between the
% wave-packets is lost, and the same holds for the quantumness of the mixture of
% common-cause and cause-effect,  

 Comparing Figs.~\ref{fig:coh2probq}(a) and \ref{fig:coh2probq}(d), the indicator of a coherent
 mixture falls more quickly than the indicator of physical mixture, giving rise to
 a range of values of the delay parameter for which the causal map is still
 a physical mixture of quantum common-cause and cause-effect mechanisms
 (\textsc{PhysQ}) but $\mathcal{N}^b_{CD}$ is zero for both $\ket{H}$ and $\ket{V}$
 on $B$. For the family of causal maps that we target in this transition, the measurement
 $\{\ket{H},\ket{V}\}$ on $B$ maximizes the induced negativity, and consequently
 finding $\mathcal{N}^b_{CD}=0\;\forall b=H,V$ suggests that no measurement
 could have induced entanglement. In this case, the causal map is an element of \textsc{PhysQ}, but not of  \textsc{Coh}.%, even if it is still in \textsc{PhysQ}.

\begin{figure*}[t] 
  \centering
  \includegraphics[scale=0.6]{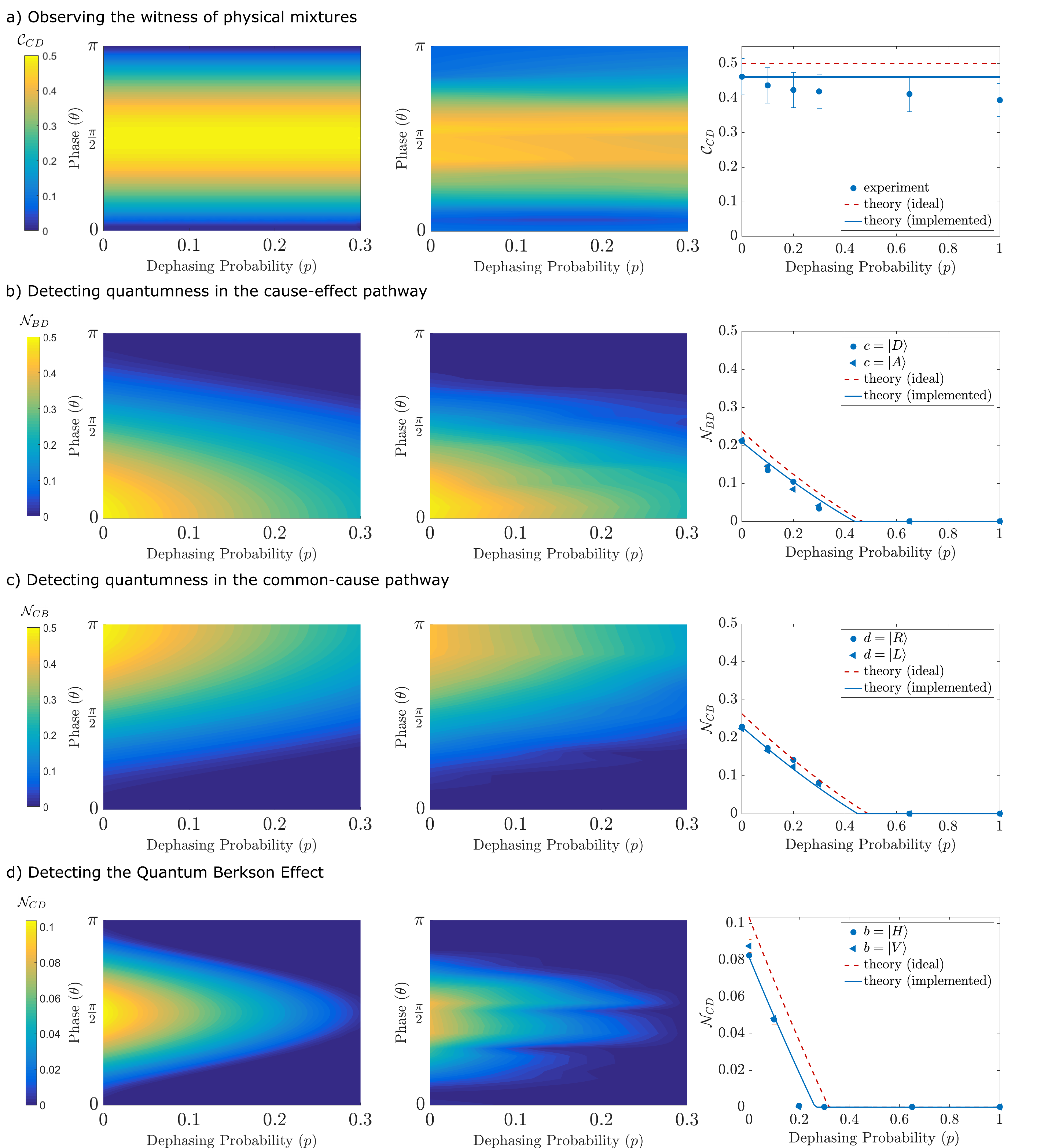}
  \caption{\footnotesize{\textbf{Classifying a family of causal relations: transition
  from Coh to PhysC.}
 % In each scenario, from top to bottom, we present:   (a) the 
 (a) Witness of physical
  mixture $\mathcal{C}_{CD}$, (b) negativity $\mathcal{N}_{BD}$ for 
  measurements of $\sigma_X$ eigenstates on $C$, (c) negativity
  $\mathcal{N}_{CB}$ for preparations of $\sigma_Y$ eigenstates on $D$, and
  (d) negativity $\mathcal{N}_{CD}$ for measurements of $\sigma_Z$ eigenstates
  on $B$. 
  %The particular preperations and measurements are chosen to maximize the induced negativity  in each case. 
  For each witness, we present: % the following information:  
  (left) theoretical predictions, which are identical for selection on either
  eigenstate;
  (middle) experimental data under selection on the $+1$ eigenstate; 
  %where, for reference, the white dashed lines indicate the 0 value contour line for the ideal causal map,
  (right) a two-dimensional cross section through the contour plot at $\theta=92.8^{\circ}$,
  comparing experimental data for both selections (blue circles and triangles) with theoretical expectations
  assuming no experimental imperfections (red curve) and with theoretical expectations given an estimate of the experimental
  imperfections inferred from the tomographically-reconstructed causal map realized
  at $p=0$ (blue curve). We observe: (a) The witness of physical mixture remains constant
  throughout the transition. % from Coh to PhysC. 
%As the dephasing probability is increased beyond $p=0.3$, the profiles for both the witness of physical mixture and the negativity remain constant.
 (b)--(d) %In the transition from Coh to PhysC, 
 Quantumness in the common-cause and cause-effect pathways persists longer than
 quantumness in the way that the two mechanisms are combined.
  %When the damping of the coherence terms is between
  %... and ..., we realize a causal map that belongs to PhysQ but not Coh. (b,c) and
 As we range from purely common-cause ($\theta=\pi$) to purely cause-effect
  ($\theta=0$), the signature of coherence in the cause-effect pathway increases,
  whereas the signature of coherence in the common-cause pathway diminishes.}}
\label{fig:coh2physc}
\end{figure*}

\begin{figure}[t]
  \includegraphics{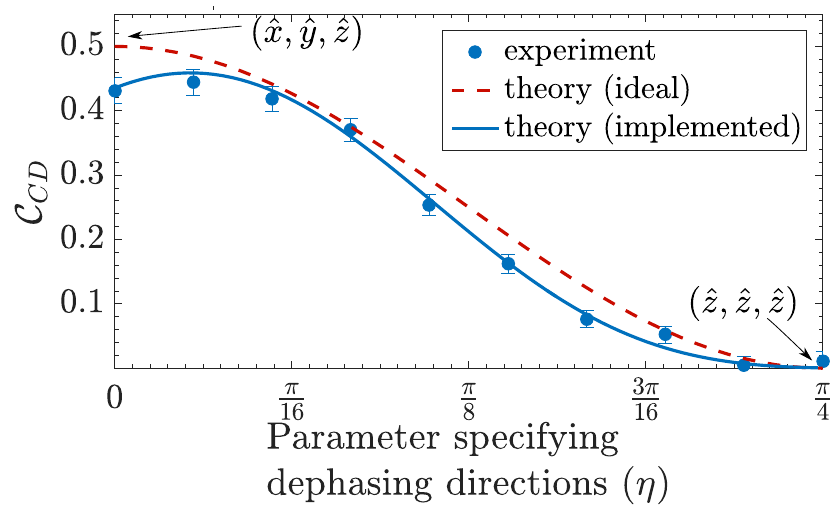} 
  \caption{\footnotesize{\textbf{Transition from PhysC to ProbC.}
  A nonzero value of the witness $\mathcal{C}_{CD}$ heralds a physical mixture of
  causal structures.  The witness is evaluated for different dephasing
  directions parametrized by Eq.~\ref{eq:eta},
  %$(\hat n_E,\hat n_D,\hat n_B)= (\cos{2\eta}~\hat  x+\sin{2\eta}~\hat z, ~\cos{2\eta}\sin 2\eta~\hat x-\cos 2\eta~\hat  y+\sin^2{2\eta}~\hat z,~\hat z$)  
  with the extremes, $\eta=0$ %($\hat x, \hat y,  \hat z$) 
  and $\eta=\pi/4$, %($\hat z,\hat z, \hat z$), 
  corresponding to settings for
  our examples of \textsc{PhysC} and \textsc{ProbC}, respectively.  For $\eta=\pi/4$,
  we measure $\mathcal{C}_{CD}=0.01\pm0.02$, which is compatible with a probabilistic
  mixture, whereas $\eta=0$ produces $\mathcal{C}_{CD}=0.44 \pm 0.02$, demonstrating
  with high confidence a non-trivial physical mixture of cause-effect and
  common-cause mechanisms.  The solid lines indicate theoretical predictions 
  assuming no experimental imperfections (red) and assuming the
  intial experimental implementation and dephasing operation inferred by comparing
  the causal maps with and without dephasing (blue).}}
 \label{fig:rotated witness}
 \end{figure}

 The indicators we observe are generally degraded by imperfections in the
 state preparation and the partial swap gate.  In order to take these into account in
 the theoretical predictions,  we use the tomographic reconstruction of the causal
 map \cite{MacLeanEtAl_2016} in the scenario with no delay to predict the witness values when $\tau$ is non-zero, yielding the blue curves in Fig.~\ref{fig:coh2probq}.
 % under total dephasing for all $\eta$.
 Similar derivations are used for the blue lines in
 Figs.~\ref{fig:coh2physc} and \ref{fig:rotated witness}, starting from the case with
 no dephasing to generate predictions for arbitrary $p$ and $\eta$.
% More details are provided in the supplementary material. 
% {\color{red}[Can we get away without supplementary material for this paper?]} 

%part 2: coh to probQ

%part 3: coh to physC

 The transition from \textsc{Coh} to \textsc{PhysC} is shown in
 Fig.~\ref{fig:coh2physc}.  The vertical axis denotes the experimentally-implemented
 value of the parameter $\theta$ in the family of partial swap unitaries described in
 Eq.~\eqref{eq:partialswap}.  This ranges from $\theta=0$ (purely
 cause-effect) to $\theta=\pi$ (purely common-cause) through a family of
 intrinsically quantum combinations of the two. The value of $\theta$ that is
 realized by the interferometer is estimated by fitting our experimental data to
 a family of causal maps where $\theta$ is the only free parameter.  The horizontal
 axis denotes the values of the dephasing probability $p$ common to the three
 dephasing operations along the fixed axes $(\hat n_E,\hat n_D,\hat n_B)=(\hat x,\hat
 y, \hat z)$.  The dephasing probability $p$ ranges from $0$ (intrinsically quantum)
 to $1$ (classical), although we display only the interesting region below $p=0.3$.
 %along the fixed axes $(\hat n_E,\hat n_D,\hat n_B)=(\hat x,\hat y, \hat z)$ with dephasing probability $p$ increasing from $0$ (intrinsically quantum) to $1$ (classical). 

%  \color{cyan} [Describe what quantity each of the four parts of the figure display:, which witnesses the presence of a physical mixture, etcetera.  Also, state here, in   the main text, that the theoretical prediction is on the left while the data is on the right.]

  Figure~\ref{fig:coh2physc}(a) depicts the value of $\mathcal{C}_{CD}$ from
  Eq.~\eqref{eq:CCD}, which witnesses the presence of a physical mixture.
  Figures~\ref{fig:coh2physc}(b)-\ref{fig:coh2physc}(d) show the value of the
  negativities [defined in Eq.~\eqref{eq:negativity}] for certain bipartite
  states.  Figure~\ref{fig:coh2physc}(b) displays $\mathcal{N}^c_{BD}$, the
  negativity of the state on $BD$ inferred from obtaining the $c$ outcome in a
  measurement of $\sigma_X$ on $C$.  Figure~\ref{fig:coh2physc}(c) displays
  $\mathcal{N}^d_{CB}$, the negativity of the state on $CB$ that results from
  the preparation of the $d$ eigenstate of $\sigma_Y$ on $D$.
  Figure~\ref{fig:coh2physc}(d) displays $\mathcal{N}^b_{CD}$, the negativity of
  the state on $CD$ inferred from obtaining the $b$ outcome in a measurement of
  $\sigma_Z$ on $B$.  The first two
  [Figs.~\ref{fig:coh2physc}(b)-\ref{fig:coh2physc}(c)] detect the presence of coherence in
  the cause-effect and common-cause pathways, respectively, while the third
  Fig.~\ref{fig:coh2physc}(d) detects the quantum Berkson effect. For each
  witness, we compare the theoretical predictions (left panel) with the
  experimental data (middle column). For the negativities
  [Fig.~\ref{fig:coh2physc}(b)--\ref{fig:coh2physc}(d)], we only show data
  obtained from preselection and postselection of the +1 eigenstate; the theoretical
  predictions are the same for both eigenstates. 

 When there is no dephasing ($p=0$), as we vary the phase from a purely common-cause
 ($\theta=0$) to a purely cause-effect ($\theta=\pi$) relation, we observe that the witness
 $\mathcal{C}_{CD}$ [Fig.~\ref{fig:coh2physc}(a)] is non-zero, indicating that we
 continue to realize a physical mixture, for all but $\theta=0,\pi$ (purely
 cause-effect and common-cause). The witness $\mathcal{N}_{BD}$
 [Fig.~\ref{fig:coh2physc}(b)] increases from zero and reaches a maximum at a purely
 cause-effect relation ($\theta=0$), whereas the witness $\mathcal{N}_{CB}$
 [Fig.~\ref{fig:coh2physc}(c)] decreases to zero from a maximum at a purely
 common-cause relation ($\theta=\pi$). Finally, $\mathcal{N}_{CD}$
 [Fig.~\ref{fig:coh2physc}(d)] reaches a maximum for an equally weighted coherent
 mixture of cause-effect and common-cause ($\theta=\pi/2$).
 This behavior is precisely what one expects theoretically 
 for a family of causal  maps, which coherently interpolate between cause-effect and
 common-cause relations.

 The dephasing bases were chosen so as to ensure that the witness of physical mixture
 remains unaffected by the dephasing.  This is possible because $\mathcal{C}_{CD}$ is
 evaluated using measurements in a single basis on each system, which implies that
 dephasing in this set of bases leaves all relevant measurement outcomes unchanged.
 As such, when the probability of dephasing, $p$, is increased, $\mathcal{C}_{CD}$
 remains constant, whereas all other witnesses are decreased.

 When the gate produces a purely cause-effect relation ($\theta=0$) and there is no
 dephasing ($p=0$), $\mathcal{N}_{BD}$ is found to be nonzero, heralding 
 quantumness in the cause-effect pathway. As  the phase or the dephasing is increased, 
 i.e., as $\theta\rightarrow \pi$ or $p\rightarrow 1$, the value of the witness decreases
 until quantumness is no longer detected.
 Likewise, $\mathcal{N}_{CB}$ detects quantumness in the
 common-cause pathway when the gate produces a purely common-cause relation
 ($\theta=\pi$) and there is no dephasing ($p=1$). As the phase is decreased or
 the dephasing increased, i.e., as $\theta\rightarrow0$ or $p\rightarrow 1$, the value of the
 witness decreases until no signature of quantumness in the common-cause pathway is
 left.
% \color{cyan} $\theta = 0$ (for all $p$) and at $p=1$ (for all $\theta$), \color{black}
 %as  $\theta \rightarrow 0$ and $p\rightarrow 1$. \color{cyan} 
%
 Finally, the witness $\mathcal{N}_{CD}$ reveals signatures of the quantum Berkson
 effect when $p=0$ except near the extremes $\theta=0,\pi$. However, as the dephasing is
 increased towards $p=1$, these signatures vanish rapidly.
% Together with the witnesses of quantumness in the individual pathways,

In the regions of the parameter space where all witnesses are nonzero, the experiment realizes a member of the class \textsc{Coh}, a quantum-coherent mixture of common-cause and cause-effect relations. 

 The right panel in Fig.~\ref{fig:coh2physc} shows two-dimensional cross-sections at
 %the value of %$\theta$ %of (b) % at the dashed
 %lines. The lines mark the value of $\theta$ 
% which, among the values realized in the experiment, came closest to a balanced partial swap, namely
 $\theta=92.8^\circ$. 
%The witness of physical mixture is nonzero for all $p$, with $\mathcal{C}_{CD}=0.46 \pm 0.04$ at $p=0$, heralding a physical mixture. This holds independently of the dephasing probability $p$, due to the particular choice of dephasing directions.
 Comparing Fig.\ref{fig:coh2physc}(d) to
 Figs.\ref{fig:coh2physc}(b) and \ref{fig:coh2physc}(c), we note that, as the dephasing increases, quantumness
 in the two pathways 
 (revealed by %the signature of which is 
 $\mathcal{N}_{CB}\ne 0$ and $\mathcal{N}_{BD}\ne 0$)
 %(measured by $\mathcal{N}_{CB}$ and $\mathcal{N}_{BD}$)
 persists for longer than the quantum Berkson effect 
  (the signature of which is $\mathcal{N}_{CD}\ne 0$).
  For the family of causal maps that we target, measuring $B$ in
 the $\{\ket{H},\ket{V}\}$ basis induces the most entanglement, and as such, 
 %for this measurement, 
 when $\mathcal{N}^b_{CD}=0\;\forall b\in \{H,V\}$, %is within error of zero,  % implies there exists 
 %since no other measurement could induce non-zero entanglement, 
 the causal map is no longer in \textsc{Coh}.
 %and consequently that the causal map is no longer a quantum-coherent mixture but belongs to the class \textsc{PhysC}. 
 Thus, we first observe a loss of coherence in the way the
 different causal pathways are combined, while remaining a physical mixture:
 a transition from \textsc{Coh} to \textsc{PhysQ}. 
 %At the same time, the noise considered in this transition also introduces 
 This is followed by a gradual loss of coherence in the cause-effect and common-cause
 pathways, leading towards the class \textsc{PhysC}.

 Finally, we explore the transition between \textsc{PhysC} and \textsc{ProbC}, which
 allows us to verify the sensitivity of the witness $\mathcal{C}_{CD}$.
% and test that it behaves as expected in the classical limit by  ranging over a transition between the paradigm examples of \textsc{PhysC} to \textsc{ProbQ}.
 Figure~\ref{fig:rotated witness} shows experimental points alongside the ideal
 theoretical curves (red) as well as theory curves based on the experimentally
 reconstructed causal map with no dephasing.  The small discrepancy between the data
 and the ideal theory (red) is likely due to a slight rotation of the LCR at $C$.
 The effects of this rotation was measured by comparing the reconstructed causal maps
 for $\theta=\pi/2$ and $\eta=0$ with ($p=1$) and without ($p=0$) dephasing, and when
 accounted for, gives the adjusted theory curve (blue).  The witness
 $\mathcal{C}_{CD}$ is clearly non-zero throughout most of the transition, thereby
 attesting to the existence of a physical mixture.  The last two data points are
 within error of zero, with $\mathcal{C}_{CD}=0.01\pm0.02$ when $\hat n_{D,E,B}=\hat
 z$ ($\eta=\pi/4$).  This is consistent with the fact that only these values of dephasing in the
 system are compatible with a probabilistic mixture. 

\section{Conclusions}

 Starting from a causal map in the class \textsc{Coh}, a quantum-coherent mixture of
 cause-effect and common-cause mechanisms, we have experimentally observed both the
 transition to a causal map in the class  \textsc{ProbQ} and the transition to
 a causal map in the class  \textsc{PhysC}.  Our results illustrate that general
 causal relations can exhibit two different types of coherence, which can vary
 independently.  We observe coherence in pathways -- that is, the quantum channel
 realizing a cause-effect relation and the bipartite state encoding a common-cause
 relation -- which is gradually lost in the transition towards \textsc{PhysC}.  We
 also observe coherence in the way that the two causal mechanisms are combined, that
 is, coherence in the mixture, which we gradually decrease in the transition towards
 \textsc{ProbQ}.  

 While the consequences of decoherence for entangled quantum states are well
 understood, for instance the loss of entanglement due to coupling to local
 independent environments \cite{almeida_2007}, a theory of decoherence for quantum
 causal relations is more subtle.  As shown here, it has an added level of richness
 due to the possibility of coherence not only in individual causal pathways, but in
 the way that these pathways are mixed.  We find that in the presence of local
 environmental noise, the quantum to classical transition for the causal pathways,
 which corresponds to decoherence on a bipartite quantum state for the common-cause
 pathway and decoherence on a channel for the cause-effect pathway, is inherently
 different from the quantum to classical transition for the way in which the pathways
 are mixed. We have determined the amount of local noise that the individual pathways
 can tolerate before losing the quantumness present in their mixture, as measured by
 the quantum Berkson effect. Moreover, we observe that the coherence in the mixture
 is more sensitive to the type of noise presently studied, decaying faster than the
 coherence in the individual causal pathways. 
  
 The experimental exploration of the transition between quantum and classical causal
 relations is based on the formalism of causal maps and classes developed in
 Refs.~\cite{RiedEtAl_2015, MacLeanEtAl_2016}, and is part of the larger program of
 advancing the causal understanding of quantum mechanics.  In particular, to explore
 the novel types of causal relations that can hold in higher dimensions and among
 larger numbers of systems, it will be necessary to rely on different types of
 controlled noise as a tool for transitioning between various classes of causal
 relations. 

%ACKNOWLEDGMENTS
\section*{Acknowledgments} 
 This research was supported in part by the Foundational Questions Institute (FQXI),
 the Natural Sciences and Engineering Research Council of Canada (NSERC), Canada
 Research Chairs, Industry Canada and the Canada Foundation for Innovation (CFI).
 Research at Perimeter Institute is supported by the Government of Canada through
 Industry Canada and by the Province of Ontario through the Ministry of Research and
 Innovation. KR would like to acknowledge the funding from the Austrian Science
 Fund (FWF) through Grant No. SFB FoQuS F4012.

 % {\bf Author contributions}
 % JPM, KR, RWS and KJR conceived the original idea for the project.  
 % KR and RWS developed the project and the theory.  
 % JPM and KJR designed the experiment.
 % JPM performed the experiment and the numerical calculations.  
 % JPM, KR, RWS and KJR analyzed the results.  
 % JPM and KR wrote the first draft of the paper and all authors contributed to the final version.

 %\bibliography{../CoherentCausation2,../causalityexp2}
%

\end{document}